\documentclass{elsart}
\usepackage{amsmath,amsfonts,amssymb}

\begin{document}

\begin{frontmatter}
\title{A Revised Generalized Kolmogorov-Sinai-like Entropy and Markov Shifts}

\author{Qiang Liu}
\address{Center for Nonlinear Complex Systems, Department of Physics, Yunnan
University, Kunming, Yunnan 650091, P. R. China} 
\author{Shou-Li Peng\corauthref{cor1}}
\address{CCAST (World Laboratory), P. O. Box 8730, Beijing 100080, P. R. China}
\address{Center for Nonlinear Complex Systems, Department of Physics, Yunnan
University, Kunming, Yunnan 650091, P. R. China}
\corauth[cor1]{Corresponding author}
\ead{slpeng@ynu.edu.cn}

\begin{abstract}
The Kolmogorov-Sinai entropy in the sense of Tsallis under Bernoulli shifts was obtained by
Mes\'{o}n and Vericat [J. Math. Phys. 37, 4480(1996)]. In this paper, we propose a revised
generalized Kolmogorov-Sinai-q entropy under Markov shifts. The form of this generalized entropy
with factor $q$ is nonextensive. The new generalized entropy contains the classical
Kolmogorov-Sinai entropy and Reny\'{i} entropy as well as Bernoulli shifts as special cases.
Applying the generalized entropy we discuss its approximate behavior qualitatively, the entropy
rate and the sensitive value $q^{*}$ of the nonextensive parameter $q$, which may exit in the
interval (-2,2) nearby where the generalized entropy return to the classical K-S entropy.
\end{abstract}

\begin{keyword}
generalized Kolmogorov-Sinai entropy, Markov shift, ergodic theory
\PACS 05.45.-a, 05.20.-y,
65.40.Gr, 02.50.Ga
\end{keyword}
\end{frontmatter}

\section{Introduction}

Both entropy theories of Boltzmann-Gibbs (BG) and Shannon are two of the most important foundations
of statistical physics and modern information theory respectively. the Kolmogorov-Sinai (KS)
entropy\cite{Sinai-1959,Sinai-1964} in the dynamic systems make a great progress, which establishes
a milestone to modern dynamic systems and affects the development of mathematical physics since
then. Many concepts characterized complex multifractal systems such as dimension, dimension
spectrum\cite{Kadanoff}, entropy, generalized entropy\cite{Grassberger1,Grassberger2}(such as
correlation entropy, particularly, Renyi entropy\cite{renyi}), Lyapunov exponent, topological
pressure\cite{Bohr}, etc., have gained very important applications in physics. More recently, a
generalized entropy to Boltzmann entropy in physical systems is proposed by
Tsallis\cite{Tsallis_1988}, which has the nonextensive property and hence may be a good description
for some complex systems with multifractal properties. The Tsallis entropy has been applied to
various physical phenomena which may deviate from equilibrium state for instance
\cite{Saslaw,Risken,Cyaceres,Montroll,Sciama,Bahcall} to particularly edge phenomena of chaotic
threshold in one-dimensional dissipative systems\cite{PRL2002,PRE2004,PRL2004}. It is worthy to
concern the progress of works in two branches to physics at present. On the one hand, the
nonextensive forms of the Tsallis entropy in some important branches of mathematics have been
developed. A generalized entropy in the Tsallis sense for the dynamic systems has been presented by
Mes\'{o}n and Vericat first time\cite{meson}. The generalization does what Kolmogorov did for
Shannon entropy under the Bernoulli scheme, and obtains a mathematical parallel which is proved as
an isomorphism invariant\cite{meson,meson1}. On the another hand, the vague relation between the BG
entropy of physics and the KS entropy of dynamics has been clarified gradually by works of many
authors\cite{Latora} as research of Tsallis entropy. The KS entropy can be interpreted with a rate
of physical entropy production per unit time. Thus the position of KS entropy of dynamics in the
application of statistical physics becomes more important.

The Mes\'{o}n-Vericat's definition of KSq entropy is nice to consider the property of power
distribution which is a common character of objects described with Tsallis entropy. Meanwhile that
Tsallis entropy can smoothly recover to BG entropy is a well-known advantage. However the
Mes\'{o}n-Vericat's definition of KSq entropy seems to be lack of the advantage, i.e. it can not
recover KS entropy when $q=1$. To this aim we would revise Mes\'{o}n-Vericat's definition and make
it more complete. Secondly, the Bernoulli shift is a basic scheme of great importance in
measure-preserving dynamic systems. When systems to be studied are extended and enlarged, it is
necessary to exploit rather complex shifts beyond Bernoulli ones. So the another aim of this paper
attempt to extend the work of Mes\'{o}n and Vericat from the Bernoulli scheme to the Markov scheme,
and present a generalized mathematical form of KS entropy in the sense of Tsallis under Markov
shifts. The generalized entropy should be consistent in mathematics with both cases of the
classical KS entropy and the Kolmogorov-Sinai-like (KSq) entropy under Bernoulli shifts and our
work may be a approach to reach this goal. Finally, interrelation of both Renyi and Tsallis entropy
is interested and focused in statistical physics community now. It is one important fact in
statistical physics that obtaining same distribution under maximal entropy principle are equivalent
because they are monotonic each other. We will see also that there is similar situation in the
level of abstract dynamics.

\section{The Revised K-S-q entropy}

\label{GMS}

Now we consider the generalized Markov shift in the measure-preserving
dynamic system. Let $k\geq 2$ be a fixed integer which is cardinal of the
symbolic set $Y=\{0,1,\ldots ,k-1\}$. Let a direct product space $(X,%
\mathcal{B})=\Pi _{-\infty }^{\infty }\{0,1,\ldots ,k-1\}$ be a space of
doubly infinite, where $\mathcal{B}$ is the subset family equipped the
product $\sigma $-algebra on the space $X$. The shift transformation $%
T:X\rightarrow X$ is $T\{x_{n}\}=\{x_{n+1}\}$ For each natural number $n$
the cylinder set
\begin{equation}
\{(x_{i})_{-\infty }^{\infty }|\text{ }x_{j}=i_{j},\text{ for }|j|\text{ }%
\leq n\}\text{ \ \ \ }(i_{j}\in Y,)
\end{equation}%
forms the basic set family with the product $\sigma $-algebra. We can
introduce a unique probability measure $m$ on $(X,\mathcal{B})$ as
\begin{equation}
m(\{x_{i}\in X|x_{l}=i_{0},x_{l+1}=i_{1},\ldots
,x_{l+n}=i_{n-1}\})=p_{n}(i_{0},i_{1},\ldots ,i_{n-1}),  \label{measure}
\end{equation}%
where probability $p_{n}(i_{0},i_{1},\ldots ,i_{n})$ is a positive real number. According to
Kolmogorov consistency theorem\cite{walters} the probability measure $m$ in (\ref{measure}) can be
extended to the entire space $(X, \mathcal{B})$ . The Markov shift is a quadruple $(X,\mathcal{B}
,m,T$) and one of the important cases of measure-preserving dynamic systems. If we may endow some
explanations to it, $X$ is regarded as a sample space made up of the whole results of individual
random experiments. Then the product $\sigma $-algebra $\mathcal{B}$ on $X$ and the probability
measure $m $ associated with the cylinder set are the standard frame for the random experiments.
For one individual experiment the sample leads to the probability of the cylinder
$p_{n}(i_{0},i_{1},\ldots ,i_{n})$. Different from the Bernoulli scheme, if we suppose that the
outcomes of every experiment are not independent, the transition probability $P_{ij}\geq 0$ between
both states $(i,j)$ depends only on the states $(i,j)$. Then we have an irreducible non-negative
$k\times k$ stochastic matrix $P=\{P_{ij}\}$ $ (P_{ij}\geq 0,i,j\in Y)$ with its row normalizing
condition $ \Sigma_{j=0}^{k-1}P_{ij}=1$ for each $i,j$ and there is a unique positive left vector
$p=$ $(p_{0},\ldots,p_{k-1})$ which satisfy the invariance probability
\begin{equation}
p_{j}=\Sigma_{i=0}^{k-1}p_{i}P_{ij}.  \label{R2}
\end{equation}%
We will assign the transition with Markov property to the probability of the
cylinder as
\begin{equation}
p_{n}(i_{0},i_{1},\ldots
,i_{n-1})=p_{i_{0}}P_{i_{0}i_{1}}P_{i_{1}i_{2}}\ldots P_{i_{n-2}i_{n-1}},
\label{2star}
\end{equation}%
with the initial probability vector i.e. $p_{0}(i_{0})=(p_{0},p_{1},\ldots
,p_{k-1})$ we obtain a \emph{two-sided $(\mathbf{p},P)$ Markov shift}. The
measure preserving is guaranteed by two normalizing conditions: $%
\Sigma_{j=0}^{k-1}P_{ij}=1$ and $\Sigma_{i_{0}\in Y}p_{0}(i_{0})=1. $ If the
transition probability is independent of the initial state $i$, i.e., taking
$P_{i,j}=p_{j}$, then the probability of cylinder (\ref{2star}) becomes
\begin{equation}
p_{n}(i_{0},i_{1},\ldots ,i_{n-1})=p_{i_{0}}p_{i_{1}}p_{i_{2}}\ldots
p_{i_{n-1}}.  \label{3star}
\end{equation}%
The \emph{$\mathbf{p}$-Bernoulli shift} is a special case of the \emph{$(%
\mathbf{p},P)$ }Markov shift.

According to Shannon's information theory\cite{Shannon}, if we already know
a point $x\in X$ belongs to some fixed set of partition $\mathcal{A}%
=\{A_{0},A_{1},\ldots ,A_{m-1}\}$, the quantity of information is
\begin{equation}
H_{1}(\mathcal{A})=-\sum_{i=0}^{m-1}\mu (A_{i})\log \mu (A_{i}),
\end{equation}%
where partition $\mathcal{A}$ should satisfies the normalizing condition $%
\sum_{i=0}^{m-1}\mu (A_{i})=1.$ This entropy possess additivity. Tsallis
presented a generalization entropy\cite{Tsallis_1988} of the physics entropy
as
\begin{equation}
H_{q}(\mathcal{A})=(q-1)^{-1}\Big(1-\sum_{i=0}^{m-1}[\mu (A_{i})]^{q}\Big),
\end{equation}%
where $q$ is an arbitrary real number, and the entropy has pseudo-additivity
\begin{equation}
H_{q}(A+B)=H_{q}(A)+H_{q}(B)+(1-q)H_{q}(A)H_{q}(B)  \label{pseudo}
\end{equation}%
when two independent subsystems $A$ and $B$ form one system. The Tsallis
entropy recovers the Shannon entropy when $q=1$. We now define a generalized
entropy parallel to Ref.\ \cite{meson} for the finite partition partition $%
\mathcal{A}=\{A_{0},A_{1},\ldots ,A_{m-1}\}$ of the sample space. Based on
Meson \& Vericat's definition\cite{meson}, a new revised definition of
generalized entropy(\thinspace KSq entropy) of the measure-preserving
transformation $T$ is
\begin{equation}
h_{KSq}(T)=(q-1)^{-1}[1-\exp [(1-q)\widetilde{h}_{KSq}(T)]],  \label{a}
\end{equation}%
\begin{equation}
\widetilde{h}_{KSq}(T)=\sup_{\mathcal{A\subseteq B},\mathcal{A}\,\text{finite%
}}h_{KSq}(\mathcal{A},T),  \label{b}
\end{equation}%
\begin{equation}
h_{KSq}(\mathcal{A},T)=\lim_{n\rightarrow \infty }\frac{1}{n}%
\{(1-q)^{-1}\log [1+(1-q)H_{q}(\bigvee_{i=0}^{n-1}T^{-i}\mathcal{A})]\}.
\label{c}
\end{equation}%
where the supremum in (\ref{b}) is taken over the set $\mathcal{B}$ formed with all finite
sub-$\sigma $-algebras $\mathcal{A}$. It is enough to physics, if not, just taking over all
countable partitions like current monographs\cite {Sinai3}. After getting rid of an infinite point
partition entropy we can still work on the finite entropy, because the finite entropy rate is a
necessary condition to statistical mechanics. Note that the partition entropy $H_{q}$ in (\ref{c})
takes the form of Tsallis entropy. The revised definition improves the nonextensive property of the
parameter $q$. Since $q$ is any real number, the case $q=1$ will be imposed continuity. By using
L'Hospital rule the definitions of nonextensive entropy(\ref{a})(\ref{b})(\ref{c}) recovers
smoothly to that of classical K-S extensive entropy
\begin{equation}
h_{KS}(T)=h_{KS1}(T)=\widetilde{h}_{KS1}(T),  \label{a1}
\end{equation}%
\begin{equation}
\widetilde{h}_{KS1}(T)=\sup_{\mathcal{A\subseteq B},\mathcal{A}\,\text{finite%
}}h_{KS1}(\mathcal{A},T),  \label{b1-1}
\end{equation}%
\begin{equation}
h_{KS1}(\mathcal{A},T)=\lim_{n\rightarrow \infty }\frac{1}{n}%
H_{1}(\bigvee_{i=-n}^{n}T^{i}\mathcal{A}).  \label{c1}
\end{equation}%
Thus the classical definition of KS\ entropy is contained within the the
nonextensive parameter $q$, so the revision enhance a little completeness to
the definition of Mes\'{o}n and Vericat\cite{meson}. We will see that some
conclusions of classical KS entropy should also be contained in the
generalized KSq entropy below.

\section{the K-S-q entropy of Markov Shifts}

We now come to seek the form of a generalized Markov shift.  There exist many generalized Markov
shifts on the basis of the classical Markov shift, we want to seek such a generalized Markov shift
which can generalize KS entropy according to the form of Tsallis entropy. We assume that there is a
family of pseudo-stochastic matrices with the real parameter $q$: $\{P_{ij}^{q}\}_{i,j\in Y},\
P_{ij}\geq 0$ and still satisfies the condition $\Sigma _{j=0}^{k-1}P_{ij}=1$. One recognizes in
the analysis of multifractal that the Tsallis entropy is analogy to $q$-dimension in its
mathematical form\cite{Gross,Weisstein}. It is necessary to introduce a new $ L_{q}$-probability
$r_{j}$:
\begin{equation}
r_{j}=(\sum_{i=0}^{k-1}p_{i}P_{ij}^{q})^{1/q},
\end{equation}%
which will be used as the partition probability of the generalized Markov shift. Of course, the
generalized probability can play the role of $p_{j}$ in the classical Markov shift. when $q$
approaches to $1$, it has $r_{j}\rightarrow p_{j}$ and recovers to the classical Markov shift.
Particularly, if $P_{ij}=p_{j}$, then $r_{j}^{q}$ is just the partition probability of the
generalized Bernoulli shift defined by Mes\'{o}n and Vericat\cite{meson}. So the partition
probability contains the generalized Bernoulli shift as a special case. It will lead to the KSq
entropy of the Markov shift we wanted.

We now calculate the KSq entropy of generalized Markov shift. With the definition of product
$\sigma $-algebra there exits a canonical generator $\mathcal{G}=\bigvee_{i=-\infty }^{\infty
}T^{i}\mathcal{A}$. For the generalized entropy the calculation is analogy to Bernoulli scheme
\cite {meson}. Considering a typical element of the refinement
$\xi(\bigvee_{i=0}^{n-1}T^{-i}\mathcal{A)}$
\begin{eqnarray*}
\lefteqn{A_{i_{0}}\cap T^{-1}A_{i_{1}}\cap \cdots\cap T^{-n+1}A_{i_{n-1}}}\\
&&\qquad =\{\{x_{n}\}:x_{0}=i_{0},x_{1}=i_{1},\ldots,x_{n-1}=i_{n-1}\},\qquad 0\leq i_{j}\leq k-1
\end{eqnarray*}%
and using the generalized Markov shift and the $L_{q}$-probability $r_{j}$ in the previous section,
we will assign the new probability measure to the
partition $\mathcal{A}$, i.e. $\mu (A_{j})=C_{j}$, we can take the $L_{q}$%
-probability partition in (\ref{c})
\begin{eqnarray*}
\lefteqn{\mu (A_{j_{0}}\cap T^{-1}A_{j_{1}}\cap \cdots \cap T^{-(n-1)}A_{j_{n-1}})}\\
&&\qquad =r(x_{0}=j_{0})r(x_{1}=j_{1})\cdots r(x_{n-1}=j_{n-1})\\
&&\qquad =r_{j_{0}}r_{j_{1}}\ldots r_{j_{n-1}}, \qquad (0\leq j_{0},j_{1},\ldots,j_{n-1}\leq k-1).
\end{eqnarray*}%
So the generalized refinement entropy of probability space can be derived by
the probability $r_{j}$
\begin{equation}
H_{KSq}(\bigvee_{i=0}^{n-1}T^{-i}\mathcal{A}))=\frac{1}{1-q}\Big[\Big(%
\sum_{j=0}^{k-1}\sum_{i=0}^{k-1}p_{i}P_{ij}^{q}\Big)^{n}-1\Big].
\label{Formation}
\end{equation}%
Put (\ref{Formation}) into (\ref{a}), (\ref{b}) and (\ref{c}), then we have
the generalized KSq entropy of transformation $T$ under Markov shifts:
\begin{eqnarray}
h_{KSq}(T) &=&\frac{1-\sum\limits_{i=0}^{k-1}\sum%
\limits_{j=0}^{k-1}p_{i}P_{ij}^{q}}{q-1} \\
&=&-\sum_{i=0}^{k-1}\sum_{j=0}^{k-1}p_{i}P_{ij}^{q}\mathrm{ln}_{q}P_{ij}.
\label{KSq-1}
\end{eqnarray}%
where
\begin{equation}
\mathrm{ln}_{q}{x}=\frac{x^{1-q}-1}{1-q}\qquad (q\neq 1).
\end{equation}%
We see that (\ref{KSq-1}) returns to the original form of KS entropy when $%
q\rightarrow 1$
\begin{equation}
h_{KS1}(T)=-\sum_{i=0}^{k-1}\sum_{j=0}^{k-1}p_{i}P_{ij}\log P_{ij},
\label{KSq-2}
\end{equation}%
and since the Bernoulli shift is a special case of the Markov shift, when $%
P_{ij}=P_{j}$, (\ref{KSq-1}) becomes
\begin{equation}
h_{q}(T)=\frac{1-\sum\limits_{i=0}^{k-1}p_{i}^{q}}{q-1}.  \label{KSq-3}
\end{equation}%
This expression recovers the generalized entropy of the Bernoulli shift obtained by Ref.\
\cite{meson}. Thus (\ref{KSq-1}) contains classical KS entropy and generalized Bernoulli KSq
entropy. Furthermore, it is noted that for the generalized Markov scheme both the \emph{entropy of
T} and \emph{entropy of T with respect to} $\mathcal{A}$ do not equal.
\begin{equation}
h_{KSq}(T)\neq h_{KSq}(\mathcal{A},T).  \label{inequality}
\end{equation}%
Using the simple calculation with definition (\ref{c}) for $(q\neq 1)$ we
have
\begin{equation*}
h_{KSq}(\mathcal{A},T)=(1-q)^{-1}\log
[\sum_{j=0}^{k-1}\sum_{i=0}^{k-1}p_{i}P_{ij}^{q}]
\end{equation*}%
This is just the well-known Renyi entropy\cite{renyi} in the Markov scheme. The \emph{entropy of
transformation T with respect to partition} $\mathcal{A}$ is one part of the definition of KSq
entropy, i.e. The KSq entropy contains Renyi entropy as a component in context of dynamic systems.
When $h_{KSq}(\mathcal{A},T)$ is maximized by supremum through (\ref{b}) and (\ref {a}), it leads
to KSq entropy. This implies an unification of both Tsallis and Renyi entropy in KSq form,
naturally, it includes a mergence of two parameters $q$ and $\beta$ which should not be confused in
the viewpoint of current issue, because there is a great difference in the additivity of both.
Particularly, we have
\begin{equation}
\lim_{q\rightarrow 1}h_{KSq}(T)=\lim_{q\rightarrow 1}h_{KSq}(\mathcal{A},T),
\label{recov1}
\end{equation}%
it shows that the both are  generalizationes of the KS entropy along the line of abstract dynamics
although Renyi and Tsallis entropy are different in the sense of Kolmogorov-Nagumo
average\cite{Dukkipati}. For the Bernoulli scheme, as it is a special case of the Markov scheme,
$h_{KSq}(T)$ does also not necessarily agree with the entropy $h_{KSq}(\mathcal{A},T)$ by
(\ref{inequality}), even  $h_{KSq}(T)=H_{q}(p)$\cite{meson} for the Bernoulli scheme of arbitrary
$q.$ According to the revised definition (\ref {c}) we still have
$h_{KSq}(\mathcal{A},T)=(1-q)^{-1}\log [\sum_{j=0}^{k-1}p_{j}^{q}]$ to be Renyi entropy in
Bernoulli scheme, so the recover (\ref{recov1}) must also be available.

Isomorphism equivalent class of the KSq entropy in Ornstein's sense is a complex problem for Markov
shifts. We will not discuss it since it is not a complete invariant for Kolmogorov automorphism.
However, isomorphism of theKSq entropy for Bernoulli shifts will be able to discuss as same as
Mes\'{o} n and Vericat\cite{meson}.

\section{Application}

The asymptotic behavior of KSq entropy is interesting, for it ties closely with the physical
meaning of the non-extensive parameter $q$. We discuss it with Beck's frame\cite{Beck} of the
statistics of dynamic symbol sequences.
For a given map $T$ the Shannon entropy $H$%
\begin{equation*}
H(\mu,\mathcal{A},n)=-\sum_{i_{0},i_{1},\ldots,i_{n-1}=0}^{k-1}\mu (i_{0},i_{1},\ldots,i_{n-1})\ln
\mu (i_{0},i_{1},\ldots,i_{n-1}),
\end{equation*}%
where $\mu (i_{0},i_{1},\ldots,i_{n-1})=\mu (A_{i_{0}}\cap T^{-1}A_{i_{1}}\cap\cdots\cap
T^{-n+1}A_{i_{n-1}})$. The study of the asymptotic behavior of the dynamic iterate sequences
involves the entropy rate per unit time
\begin{equation*}
h(\mu ,\mathcal{A})=\lim_{n\rightarrow \infty }(H/n).
\end{equation*}%
The quantity $h(\mu ,\mathcal{A})$ depends still on the partition $\mathcal{A%
}$. Maximizing it over all possible partition set $\{\mathcal{A\}}$ in the
phase space

\begin{equation}
h(\mu )=\sup_{\{\mathcal{A\}}}h(\mu ,\{\mathcal{A\}}).
\end{equation}%
The K-S entropy of the physical version is obtained, which is a global quantity to describe chaos
solely. when we focus on the numerical calculation, the supremum can be replaced by the
maximization if the generating partition exists\cite{cornfield}. In order to clarity the vague
connection between the physical entropy and the KS entropy. Latora and Baranger\cite{Latora}
calculate the entropy rate\cite{PRL2002}
\begin{equation}
\kappa _{1}=\lim_{t\rightarrow \infty }\lim_{W\rightarrow \infty
}\lim_{N\rightarrow \infty }\frac{S_{1}(t)}{t},
\end{equation}%
where the iterate number $t=n$, $W$ is the cell number of partition of the phase space worked.
Initially putting $N$($N\gg W$) points into one cell randomly, then the occupancies $\{N_{i}(t)\}$
of all cells can calculate probabilities $p_{i}(t)=N_{i}(t)/\sum_{i=0}^{w}N_{i}(t)$ numerically. It
is
found by the numerical experiment that if the $S(t)$ is given by%
\begin{equation*}
S_{1}(t)=-\sum_{i=0}^{w}p_{i}(t)\ln p_{i}(t),
\end{equation*}%
then there exists a linear value of the entropy rate which is just the KS entropy\cite{Latora}. To
describe the edge-of-chaos, the Boltzmann-Gibbs-Shannon entropy above is not be appropriate. The
non-extensive entropy must be applied\cite{Latora2}.
\begin{equation}
\kappa _{q}=\lim_{t\rightarrow \infty }\lim_{W\rightarrow \infty
}\lim_{N\rightarrow \infty }\frac{S_{q}(t)}{t}.
\end{equation}%
For the Markov scheme here we have the non-extensive entropy to be
\begin{equation}
S_{q}(t)=\frac{1-\sum\limits_{i=0}^{w}r_{i}^{q}(t)}{q-1}.  \label{r}
\end{equation}%
The asymptotic behavior of the nonextensive entropy (\ref{r}) is monotonic decrease, as the
parameter $q$ approaches to infinity i.e. $lim_{q\to -\infty }h_{KSq}(T)\to \infty$ and $
\lim_{q\to \infty }h_{KSq}(T)\to 0$ (even $ \lim_{q\to \infty }\mathrm{ln}_{q}{x\to \infty )}.$ It
is allowed that there may exist a special sensitive value $q^{*}\neq 1$ between both entropic
values $(\infty ,0)$. The time evolution of the non-extensive entropy will sharply turn when $q$
approaches $q^{*}$, which is responsible to the power-law with complicated case in nonextensive
statistics such as edge-of-chaos and the chaos nearly by eventual periodic orbits. The latter is
another power-law example to be discussed. It is noted that the continuous at $q\to 1$ guarantees
the nonextensive entropy is smooth and applicable. Since it has recognized that the KS entropy is a
production rate of Boltzmann entropy pre unite time in physical systems, it leads to that this
production rate will decay as $q$ increases. So the case $q<1$ will have more high production rate,
it enhances the randomness of system, and the case $q>1$ decreases the randomness in the Markov
shift. Thus the parameter $q$ may be a factor which can control and influence random, irregular
chaotic behavior. For the Bernoulli case the conclusion is the same as the Markov one in
qualitative. Finally, in order to estimate $q^{*}$ we discuss another special value $q=0$. For the
the \emph{entropy of T with respect to $\{\mathcal{A}\}$}, the Renyi entropy is just the
topological entropy $h_{KS0}(\mathcal{A},T)\backsim \log k$, which is smaller than the K-S-q
entropy $h_{KS0}(T)\backsim (k-1)$, i.e., $h_{KSq}(\mathcal{A},T)\leq h_{KSq}(T)$ will always hold.
We may reasonably estimate that the sensitive value $q^{*}$ should locate in a symmetric interval
with the center at $q=0$ and the two symmetric terminal points are $(-2,2)$, i.e., $-2<q^{*}<2.$The
prediction is right for the edge-of-chaos\cite{Latora2}. To confirm the property of the generalized
KSq entropy, we have applied it to eventual periodic orbits and their edge-of-chaos with symbolic
dynamics of the unimodal transformation\cite{ours} and numerical results show us that the KSq
entropy indeed has good monotonic property with $q$ and sharply turn at the interval nearby where
it recovers to KS entropy at the limit $q\to1$.

Markov shifts are available for use of physics. Applications of the theory of Markov shifts widely
spread in various practical fields, for instance, continuous Markov processes by the time shift
operator and discrete Markov chains. Many practical applications are discussed in details by many
books, for instance in the book of Bharucha-Reid\cite{Reid}, such as the growth of populations,
mutation progress, spread of epidemics, theory of gene frequencies in biology; theory of cascading
progress in physics; theory of cosmic rays in astronomy; chemical reaction kinetics in chemistry;
theory of queues, and so on. Particularly, it is interesting in the symbolic dynamics associated
with edge of chaotic phenomena in one-dimensional dissipative systems where the topological Markov
shift on discrete chain plays an very important role.

\section{acknowledgments}

The authors thank to Prof.\ K. F. Cao for his useful discussion. The work is
supported by the National Natural Science Foundation of China(Grant No.\
10565004) and the Special Research Funds for Doctoral Programs in Colleges
and Universities in China(Grant No.\ 20050673001).

\end{document}